\documentclass[aps,twocolumn,prl,floatfix,showpacs,citeautoscript,superscriptaddress,longbibliography]{revtex4-1}

\usepackage{siunitx}
\usepackage[none]{hyphenat}
\usepackage{amssymb}
\usepackage{amsmath}
\usepackage{graphicx}
\usepackage{bm}
\usepackage{times}
\usepackage{color}
\usepackage[dvipsnames,svgnames,table]{xcolor}
\usepackage{hyperref}
\usepackage{url}
\usepackage{fancyhdr}
\usepackage{enumitem}
\usepackage[english]{babel}
\usepackage[caption=false]{subfig}
\usepackage{braket}
\usepackage{scalerel}
\usepackage{graphicx}
\usepackage{mathrsfs}
\usepackage{MnSymbol}
\usepackage{lipsum}
\usepackage{soul}
\hypersetup{
pdfstartview={FitH},
colorlinks=true,    
linkcolor=NavyBlue, 
citecolor=Blue,   
filecolor=NavyBlue, 
urlcolor=NavyBlue   
}

\renewcommand{\footnotesize}{\fontsize{8bp}{1em}\selectfont}
\makeatletter
\renewcommand\footnotesize{%
   \@setfontsize\footnotesize\@ixpt{8}%
   \abovedisplayskip 8\p@ \@plus2\p@ \@minus4\p@
   \abovedisplayshortskip \z@ \@plus\p@
   \belowdisplayshortskip 4\p@ \@plus2\p@ \@minus2\p@
   \def\@listi{\leftmargin\leftmargini
               \topsep 4\p@ \@plus2\p@ \@minus2\p@
               \parsep 2\p@ \@plus\p@ \@minus\p@
               \itemsep \parsep}%
   \belowdisplayskip \abovedisplayskip
}

\begin{document}
\title{
Coherent control of photoconductivity in graphene nanoribbons}
\author{H. P. Ojeda Collado} 
\email{These authors have contributed equally}
\affiliation{Center for Optical Quantum Technologies, University of Hamburg, Hamburg, Germany}
\affiliation{Institute for Quantum Physics, University of Hamburg, Hamburg, Germany}
\affiliation{The Hamburg Center for Ultrafast Imaging, Hamburg, Germany}
\author{Lukas Broers}
\email{These authors have contributed equally}
\affiliation{Center for Optical Quantum Technologies, University of Hamburg, Hamburg, Germany}
\affiliation{Institute for Quantum Physics, University of Hamburg, Hamburg, Germany}
\author{Ludwig Mathey}
\affiliation{Center for Optical Quantum Technologies, University of Hamburg, Hamburg, Germany}
\affiliation{Institute for Quantum Physics, University of Hamburg, Hamburg, Germany}
\affiliation{The Hamburg Center for Ultrafast Imaging, Hamburg, Germany}
\begin{abstract}
We study the photoconductivity response of graphene nanoribbons with armchair edges in the presence of dissipation using a Lindblad-von Neumann master equation formalism. We propose to control the transport properties by illuminating the system with light that is linearly polarized along the finite direction of the nanoribbon while probing along the extended direction. 
We demonstrate that the largest steady-state photocurrent occurs for a driving frequency that is slightly blue-detuned to the electronic band gap proportional to the width of the nanoribbon.
We compare the photoconductivity in the presence of coherent and incoherent light and conclude that the enhancement of the photoconductivity for blue-detuned driving relies on the coherence of the driving term.
Based on this result we propose a switching protocol for fast control of the photocurrent on a time scale of a few picoseconds. Furthermore, we suggest a design for a heterostructure of a graphene nanoribbon and a high-$T_c$ superconductor, that is operated as a transistor as a step towards next-generation coherent electronics.
\end{abstract} 
\maketitle


Advances in nonlinear optics and time-resolved spectroscopies have enabled the study of coherent elementary excitations of quantum systems via pump-probe experiments. This has led to the discovery of surprising phenomena, including light-induced superconductivity~\cite{mitrano_possible_2016,budden_evidence_2021,rowe_giant_2023,eckhardt_theory_2023,von_hoegen_amplification_2022,michael_parametric_2020}, the anomalous Hall effect~\cite{McIver2020}, and the creation of exotic phases that are only possible in the time domain, such as time crystals~\cite{Else2020,Zhang2017,Choi2017,Kessler2021,Kongkhambut2021,Taheri2022,Zaletel2023,HP_2021,HP_2023}.

In addition to the exploration of these fundamental phenomena, it is of great interest to harness the control of matter using light for technological applications. One step in this direction is the concept of Floquet engineering, where controlling and engineering material properties, including topological features, has been demonstrated in numerous platforms, such as graphene~\cite{McIver2020,Lindner2011,Silvia13,Usaj14,piskunow14,Moore15,OkaFloquet,Rudner2020,fluery,Rechtsman,Klembt}. Light-control of electronic transport in graphene has also been the subject of intense research~\cite{Efetov08,Foa12,Krist16,Hanggi91,Grifoni98,Frasca03,Gagnon16,Barata20,Zeb08,Andrii06,HP13,Jellal23,Mishchenko09,Sipe14,Sipe14b,Sipe15,Mikhailov,Suzuki18,Lukas21,Hommelhoff17,Hommelhoff22,Hommelhoff23,Jensen13,Singh24,Mohamed24}. Relevant results for applications include coherent destruction of tunneling~\cite{Hanggi91,Grifoni98,Frasca03,Gagnon16,Barata20}, photon-assisted tunneling~\cite{Zeb08,Andrii06,HP13,Jellal23}, nonlinear optical transport effects~\cite{Mishchenko09,Sipe14,Sipe14b,Sipe15,Mikhailov} and ultrafast photoconductivity~\cite{Jensen13,Singh24,Mohamed24}. More interestingly, the recent experimental demonstration of coherent control of electron dynamics in graphene~\cite{Hommelhoff17,Hommelhoff22,Hommelhoff23} could pave the way to the creation of coherent electronics, which rely on the utilization of coherent excitations.

Not only does graphene emerge as a promising candidate for coherent electronics, but also some of its variants, graphene nanoribbons and graphene nanotubes, are of immediate relevance. 
The latter offer high tunability, feasible integration into solid-state architectures and constitute versatile platforms for electronic and optoelectronic technologies~\cite{CarbonElectronics,Wang}. 
Graphene nanoribbons can be synthesized with atomic precision~\cite{graphenenanoribbonprecise} and have been proposed for room temperature transistors~\cite{graphenetransistor} and photodetectors~\cite{graphenephotodetector}  due to their high mobility properties. This has triggered several theoretical studies on periodically driven transport in graphene nanoribbons~\cite{Egger14,Riku14,Riku17,Riku19}.


\begin{figure}[tb]
    \centering 
    \includegraphics[width=\linewidth]{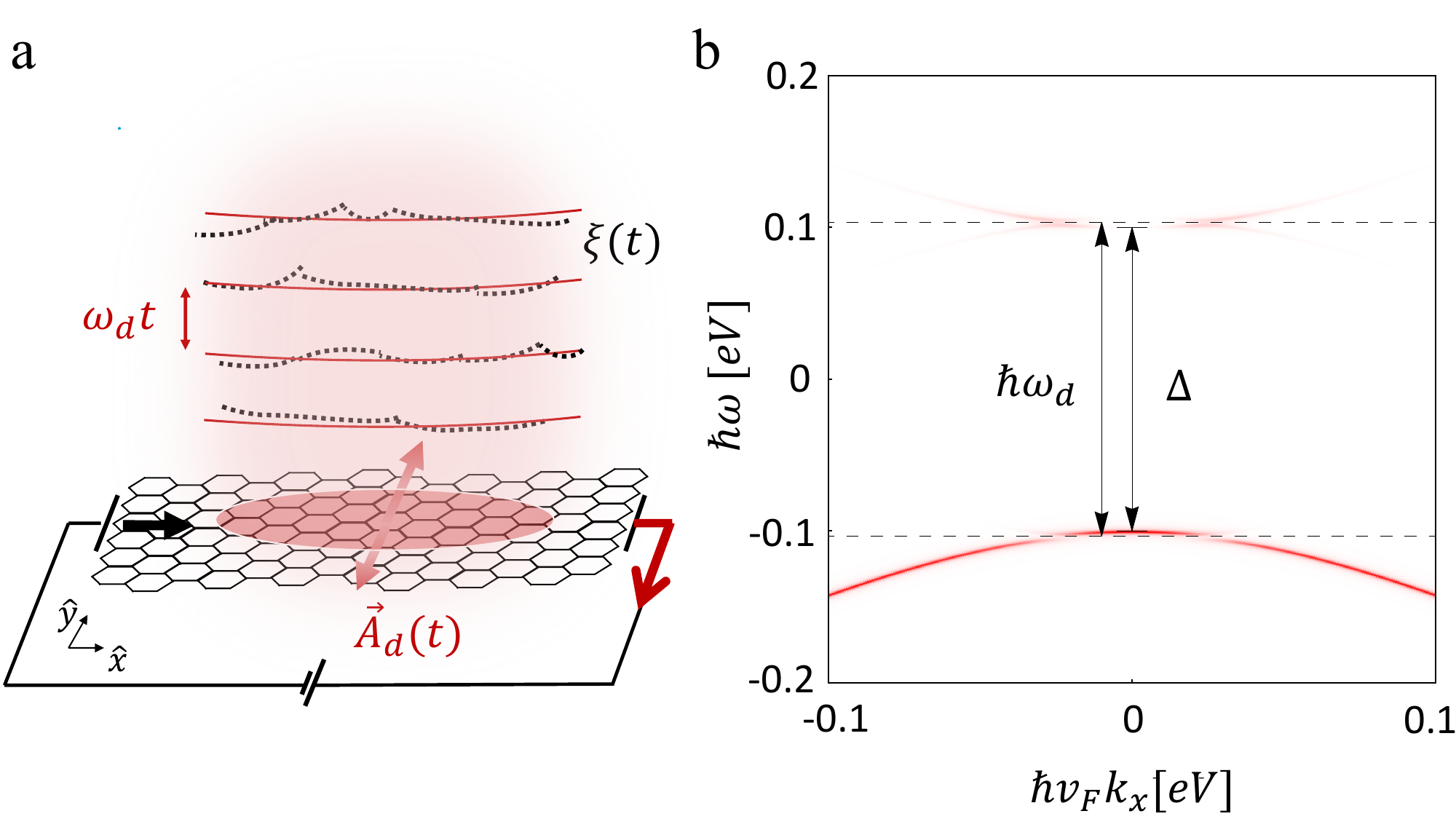}
    \caption{(a) Cartoon of a device where the current passing through a graphene nanoribbon is being controlled by a continuous coherent light (red) of frequency $\omega_{d}$ that is linearly polarized along the $\hat{y}$ direction. $\Vec{A}_d(t)$ is the vector potential. $\xi(t)$ represents a deformation in the wavefronts of light simulating an incoherent light source (black dashed lines). (b) Floquet band population (red) for an armchair nanoribbon where the coherent driving 
$\hbar\omega_{d}$ is slightly above the bare band gap $\Delta$ (marked by horizontal dashed lines) which is the optimal condition to enhance the photoconductivity.
    }
    \label{frequencyscan}
\end{figure}

In this work, inspired by recent experimental advances~\cite{Hommelhoff17,Hommelhoff22,Hommelhoff23}, we study a continuously driven graphene nanoribbon as a platform for coherent electronics. We focus on armchair-edge nanoribbons that exhibit an energy gap in the band structure depending on the width of the nanoribbon~\cite{ArmchairGNR}, allowing the gap to be tuned to match the light frequency at which one wishes to operate. Specifically, we propose a protocol in which the nanoribbon is illuminated by terahertz (THz) light linearly polarized along the finite direction and electrically probed along the extended direction of the nanoribbon using a DC bias, as we illustrate in Fig.~\ref{frequencyscan} (a). Considering relevant dissipative processes for this system, we use a Lindblad-von Neumann master equation formalism~\cite{Nuske20} and show that the steady-state longitudinal photoconductivity reaches its maximum value when the driving frequency is weakly blue-detuned to the energy gap (see Fig.~\ref{frequencyscan}(b)).
We analyze the dynamics for coherent and incoherent drives and demonstrate that the main photoconductivty features are sensitive to the temporal coherence of the driving field. In particular, we identify a large portion of the photoconductivity that is solely due to coherent phenomena that can not be captured by a semiclassical Boltzmann theory, providing an example of coherent control of electronic transport. We show the feasibility of our setup for optical photo-switches that can operate on a time scale of a few picoseconds as determined by realistic disspative processes in graphene. Based on these results, we conclude by proposing a nanoribbon-insulating-superconductor heterostructure to operate as a transistor using coherent electronics.

\section{Model and methodology}

We consider an armchair graphene nanoribbon that is driven by linearly polarized light along its finite dimension $\hat{y}$ with an associated vector potential
\begin{equation}
    \Vec{A}_d(t) = \frac{E_d}{\omega_d} \cos(\omega_d t + \xi(t)) \hat{y}.
    \label{drive}
\end{equation}
We probe the conductivity of the nanoribbon along its extended direction $\hat{x}$ with a DC bias of $E_p=E_0\hat{x}$. (see Fig.~\ref{frequencyscan} (a)). $E_d$ is the driving field strength, $\omega_d$ is the driving frequency, and $E_0$ is the probing field strength. $\xi(t)$ is a fluctuating phase that we introduce to simulate incoherent driving in order to draw a comparison to the case of coherent driving in the discussion below. In particular, we consider a phase diffusion process given by
\begin{equation}
    \xi_{t} = (1-\lambda) \xi_{t-\Delta t}  + \lambda \sqrt{\Delta t / \tau} \mathcal{W}_t
    \label{random}
\end{equation}
where $\mathcal{W}_t$ is an integrated discrete white noise series, i.e. a random walk in itself, that we take Gaussian distributed with mean $0$ and standard deviation $s$, and $\tau=2\pi/\omega_d$. This choice of phase diffusion introduces a broadening in the power spectra of the driving field which is proportional to $s^2$ for $\lambda=1$~\cite{Blachman,Middleton}. $\lambda$ is a filter parameter that allows interpolating between white noise for $\lambda=1$ and the absence of phase diffusion for $\lambda=0$. For the simulation of coherent driving we choose $\lambda=0$, i.e. $\xi_{t}$ is a constant that we consider equal to zero without loss of generality.
For incoherent driving we use $\lambda=0.0005$, which produces a smooth (strongly filtered) random walk for $\xi_{t}$ and a significant broadening in the driving field power spectra.

The system is described by the Hamiltonian $H=\sum_\mathbf{k} H_\mathbf{k} (t)=\sum_\mathbf{k} H^0_\mathbf{k}+H^d_\mathbf{k}(t)+H^p_\mathbf{k}(t)$ with
\begin{equation}
    H^0_\mathbf{k} = \Psi^{\dagger}_\mathbf{k} \left[ \hbar v_F (k_x\sigma_x+k_y\sigma_y)\right]\Psi_\mathbf{k},
    \label{hamiltnoian0}
\end{equation}
\begin{equation}
H^d_\mathbf{k}(t)=\Psi^{\dagger}_\mathbf{k}\left[e v_F A_d (t)\sigma_y\right]\Psi_\mathbf{k},
    \label{hamiltnoiand}
\end{equation}
\begin{equation}
    H^p_\mathbf{k}(t) =\Psi^{\dagger}_\mathbf{k} \left[ e v_F E_0 t 
\sigma_x\right]\Psi_\mathbf{k},
    \label{hamiltnoianp}
\end{equation}
where $v_F\approx10^6\si{\meter\per\second}$ is the Fermi velocity of graphene and $e$ is the elementary charge. Here,
$\sigma_x$ and $\sigma_y$ are embeddings of the first two Pauli matrices into four dimension. We consider a four-dimensional Hilbert space $\Psi^{\dagger}_\mathbf{k}=(|11\rangle, |01\rangle, |10\rangle, |00\rangle)$ that takes into account four possible states. $|11\rangle=\hat{c}_{\mathbf{k} B}^{\dagger} \hat{c}_{\mathbf{k} A}^{\dagger} |00\rangle$ represents a doubly occupied state with electrons on both sublattices A and B, $|01\rangle=\hat{c}_{\mathbf{k} A}^{\dagger}|00\rangle$ accounts for the state with an electron on the sublattice A, $|10\rangle=\hat{c}_{\mathbf{k} B}^{\dagger}|00\rangle$ is the state with an electron on the sublattice B, and the empty state is denoted by $|00\rangle$. See Appendix of Ref.~\cite{Nuske20} for details.

We obtain the dynamics of the system in the presence of the external drive as given Eq.~\eqref{drive}, by propagating the density matrix operator using the Lindblad-von Neumann master equation 
\begin{equation}
    \dot\rho_\mathbf{k} = \frac{i}{\hbar}[\rho_\mathbf{k},H_\mathbf{k}(t)] + \sum_{j} \gamma_j(L_j \rho_\mathbf{k} L^\dagger_j - \frac{1}{2}\{L^\dagger_jL_j,\rho_\mathbf{k}\}).
\end{equation}
The indices $j$ of the Lindblad operators $L_j$ describe the different dissipative processes of spontaneous decay, excitation, dephasing, and incoherent exchange with an electronic backgate.
The corresponding dissipation rates are $\gamma_{-}$, $\gamma_+$, $\gamma_z$ and $\gamma_\mathrm{bg}$, respectively.
We use the values $\gamma_-+\gamma_+=0.5\si{\tera\hertz}$, $\gamma_z=1.125\si{\tera\hertz}$, and $\gamma_\mathrm{bg}=1.25\si{\tera\hertz}$, which are similar to those used in previous work~\cite{Nuske20, Lukas21, Broers22}. Throughout this work, when we change the dissipation, we always do so by rescaling the coefficients by a factor $\alpha$ that keeps the ratio of these coefficients fixed, such that $\alpha=1$ corresponds to the values above. 
The temperature $T$ of the system enters the model through Boltzmann factors of conjugate processes, e.g. $\gamma_+=\gamma_-\exp\{-\frac{2\epsilon_{\bf{k}}}{k_B T}\}$, where $\epsilon_{\bf{k}}$ is the instantaneous eigenenergy scale of the driven Hamiltonian. 
The Lindblad operators $L_j$ act in the instantaneous eigenbasis of the Hamiltonian.
For further details of this method we refer to previous works \cite{Nuske20,Lukas21,Broers22}.

We focus on the dynamics of a graphene nanoribbon with armchair edges. For this case, the transverse momentum is quantized as $k_y=\frac{2\pi n}{(N+1) \sqrt{3}a}$ where $n$ is a natural number and $a\approx 1.42\si{\angstrom}$ is the lattice constant of graphene. We are interested in the low-energy physics provided by the $k_y$ channel that is closest to the Dirac points $K$ and $K'$. 
We choose the Dirac point $K = \left( \frac{4\pi}{3a}, 0 \right)$ to be the coordinate origin such that the smallest available transverse momentum is $k_y=\frac{2\pi}{3\sqrt{3}(N+1)a}$. We consider a nanoribbon of $N=55$ sites along the $\hat{y}$ direction which corresponds to a nanoribbon width of approximately $W=67 \si{\nano\meter}$ giving rise to a band gap $\Delta\sim 2\hbar v_F k_y \sim 200$meV, i.e. a frequency of about $2\pi\times 48 \si{\tera\hertz}$ (see Fig.~\ref{frequencyscan} (b)). We analyze the low temperature regime ($T=80$K) such that $\Delta \gg k_B T$ so the system behaves like a semiconductor in the absence of laser driving.

The two observables that we are interested in are the steady-state longitudinal photocurrent
\begin{equation}
j_x=\frac{n_s n_v e v_F  2\pi }{{4\pi^2} a\sqrt{3}(N+1)} \int_\mathbb{R} {\rm{d}}k_x
\frac{\omega_d}{2\pi}\int\nolimits_{t }^{t +\frac{2\pi }{{\omega }_{d}}} {\rm{d}}t' {\rm{Tr}} ({\rho }_{\bf{k}}(t'){\sigma }_{\mathrm{x}})
\label{photocurrent}
\end{equation}
and the associated longitudinal photoconductivity $G_x=j_x/E_0$.
In the above expression $t$ is a point in time at which the system has reached a steady state in the co-moving frame $k_x\xrightarrow{}k_x-eE_0 t/\hbar$, $n_{s}=2$ is the spin-degeneracy and $n_{v}=2$ is the valley-degeneracy. The photoconductivity computed in this way correspond to a local quantity that must be rescaled depending on the nanoribbon geometry by the factor $W/L$, where $W$ and $L$ are the width and length of the nanoribbon respectively.


In contrast with the full photocurrent expression  Eq.~\eqref{photocurrent} and associated photoconductivity $G_x$, we also perform calculations based on a semiclassical Boltzmann-like theory of transport in which we compute the current as $j_x=e\mathop{\sum}\limits_{\bf{k}} n_{\bf{k}} v_{\bf{k}}$, where 
\begin{equation}
n_{\bf{k}}=\frac{{\omega }_{d}}{2\pi}\int\nolimits_{t}^{t +\frac{2\pi }{{\omega }_{d}}} {\rm{Tr}} ({\rho }_{\bf{k}}(t'){H}_{\bf{k}}(t')/\epsilon_{\bf{k}})dt'
\label{photocurrent_c}
\end{equation}
is the nonequilibrium electronic distribution averaged over one period of the drive in the steady state. $v_{\bf{k}}$ is the band velocity using the instantaneous energy scale $\epsilon_{\bf{k}}$. Such an approach to compute the photocurrent does not fully consider all aspect of the dynamics. In particular coherent phenomena, such as Rabi oscillations etc., are not be captured within this treatment.

For our simulations, we consider driving field strengths up to $E_d=40\si{\mega\volt\per\meter}$ and driving frequencies in the range of up to $\omega_d=2\pi\times60\si{\tera\hertz}$, while $E_0$ is small enough compared to $E_d$ such that the DC bias is considered a linear probing field. We typically choose $E_0$ to be six orders of magnitude smaller than $E_d$.

\section{Numerical results for the Longitudinal photoconductivity}

\begin{figure}[tb]
    \centering 
    \includegraphics[width=0.9\linewidth]{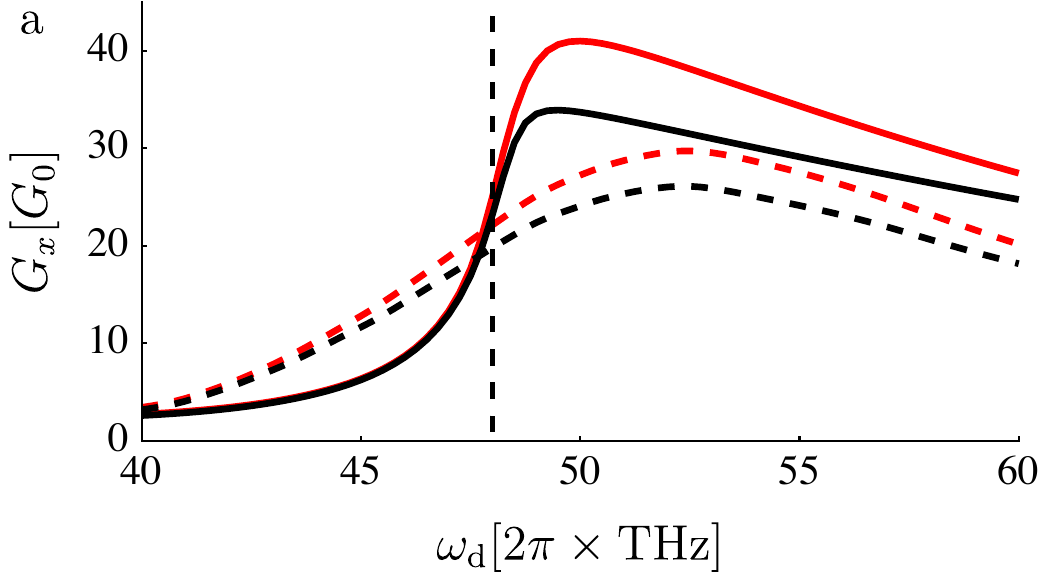}
    \includegraphics[width=0.9\linewidth]{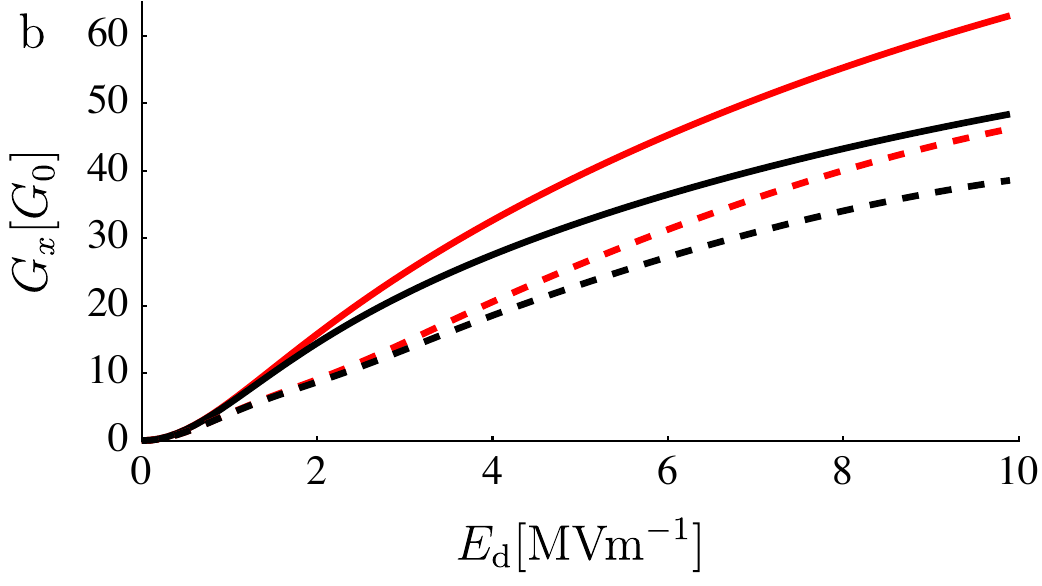}
    \includegraphics[width=0.9\linewidth]{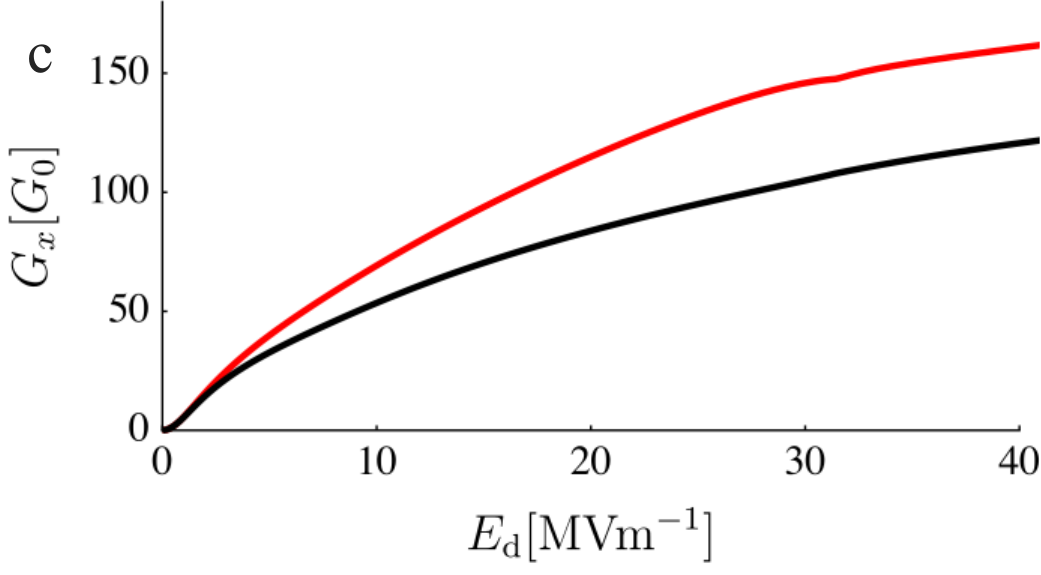}
    \caption{
       Steady-state longitudinal photoconductivity as a function of the driving frequency $\omega_d$ (a) and the driving field strength $E_d$ (b)-(c) in units of the conductance quantum $G_0=2e^2/h$. Full and dashed lines correspond to the photoconductivity produced by coherent and noisy drive, respectively. Red lines indicate photoconductivities that are evaluated by using the master equation formalism whereas the black lines correspond to a calculation using a semiclassical Boltzmann-like analysis (see main-text). In panel (a)  $E_d=5\si{\mega\volt\per\meter}$ whereas $\omega_d=\omega_d^*=2\pi\times 50$THz is used in panel (b) and (c). The vertical dashed line in panel (a) marks the electronic band gap. In panel (c) we plot the photoconductivity for larger value of driving field strength $E_d$ to show the onset of saturation. 
    }
    \label{fig2}
\end{figure}

In Fig.~\ref{fig2} we show the longitudinal steady-state photoconductivity $G_x$ for the case of perfectly coherent driving with red solid lines and that for a noisy drive with red dashed lines. As discussed above, we model coherent driving via $\lambda=0$, i.e. $\xi_t=0$, and incoherent driving via $\lambda=0.0005$, which generates a stochastic process for $\xi_t$. The black lines correspond to the semi-classical calculations of the photoconductivity.

In Fig.~\ref{fig2} (a) we show the photoconductivity $G_x$ as a function of the driving frequency $\omega_d$ for a fixed driving field strength $E_d=5\si{\mega\volt\per\meter}$. For coherent driving at frequencies below the electronic band gap $\Delta= 2\pi\times 48\si{\tera\hertz}$ (marked by vertical dashed line), the photoconductivity decreases very quickly. Even though the density of states is maximal at the bottom of the band gap, the highest conductivity occurs for slightly larger frequency  $\omega^*_d\sim 2\pi\times 50\si{\tera\hertz}$. 
By increasing the driving frequency further, the photoconductivity starts to decrease again. 
This means that by tuning the driving frequency around the band gap one can control the conductivity through the nanoribbon in a wide range between a finite maximal value for $\omega_d\sim \omega^*_d$ and very small conductivity for $\hbar\omega_d < \Delta$.

For incoherent driving (red dashed lines), we find an overall reduction and a broadening of the peak of the photoconductivity, as well as a slight shift to higher frequencies. We note that these effects become more pronounced by increasing the noise strength which is controlled by $\lambda$. This demonstrates how coherently driving the electronic system improves the degree of control over the photoconductivity compared to incoherent driving.

While the overall shape of photoconductivity as a function of $\omega_d$ holds for different values of $E_d$, its magnitude and in particular the blue-detuned photoconductivity peak at $\omega^*_d$ increases by increasing the driving field strength.
In Fig.~\ref{fig2} (b) we show the maximum value of the photoconductivity at $\omega_d=\omega^*_d$ as a function of $E_d$. For small driving field strengths it starts to increase quadratically and then changes behaviour for higher $E_d$ values. 
In the case of noisy driving (red dashed line), the photoconductivity grows more linearly as a function of $E_d$ and more importantly it is always below those photoconductivity values obtained for coherent drive (red solid line). 
These two results indicate that both the blue shift in the maximum of the photoconductivity as well as the larger values of photoconductivity reached for coherent driving are based on intrinsic coherent processes that play a key role in the charge transport.

In order to support this interpretation, we compare the full photoconductivity using the Lindblad-von Neumann master equation formalism with that computed using a semiclassical Boltzmann approach (see Eq.~\ref{photocurrent_c}) which is plotted in black in Fig.~\ref{fig2}. Solid and dashed black lines correspond to the semiclassical photoconductivity for coherent and incoherent driving, respectively.
In both cases, for small driving frequencies $\hbar\omega_d<\Delta$, the semiclassical calculation matches the full photoconductivity. However, a substantial deviation between these two descriptions becomes apparent for driving frequencies that exceed the gap, in particular around $\omega_d^*$ where the photoconductivity is maximal for coherent driving. The difference between these two approaches allows us to quantify the contribution to the photoconductivity that appears only due to coherent processes which for these parameters is about 25$\%$. This gain of the photoconductivity constitutes an example of coherent control of transport properties in graphene nanoribbons.

Note that for incoherent driving (dashed lines), the photoconductivities computed using these two different approaches show better agreement which means that a noisy drive is less efficient  in activating coherent phenomena contributing to the photocurrent. In that case, almost the entire photoconductivity can be explained by a semiclassical Boltzmann description. In Fig.~\ref{fig2}~(b), the discrepancy between the two dashed lines appears only for much larger values of $E_d$ in comparison with the coherent driving scenario. This demonstrates that stronger $E_d$ are needed in the case of noisy driving in order to produce a coherent contribution to the photocurrent. 

In Fig.~\ref{fig2} (c), we further increase the driving field strength $E_d$ up to $40\si{\mega\volt\per\meter}$. The photoconductivity changes from an approximately linear behavior to an onset of saturation in agreement with recent experimental results~\cite{Singh24}. In this regime, the coherent contribution to the photoconductivity continues to increase, as stronger driving generates more interband coherence in the system.

 \section{Graphene-based photoswitches}

\begin{figure}[b]
    \centering 
    \includegraphics[width=1.02\linewidth]{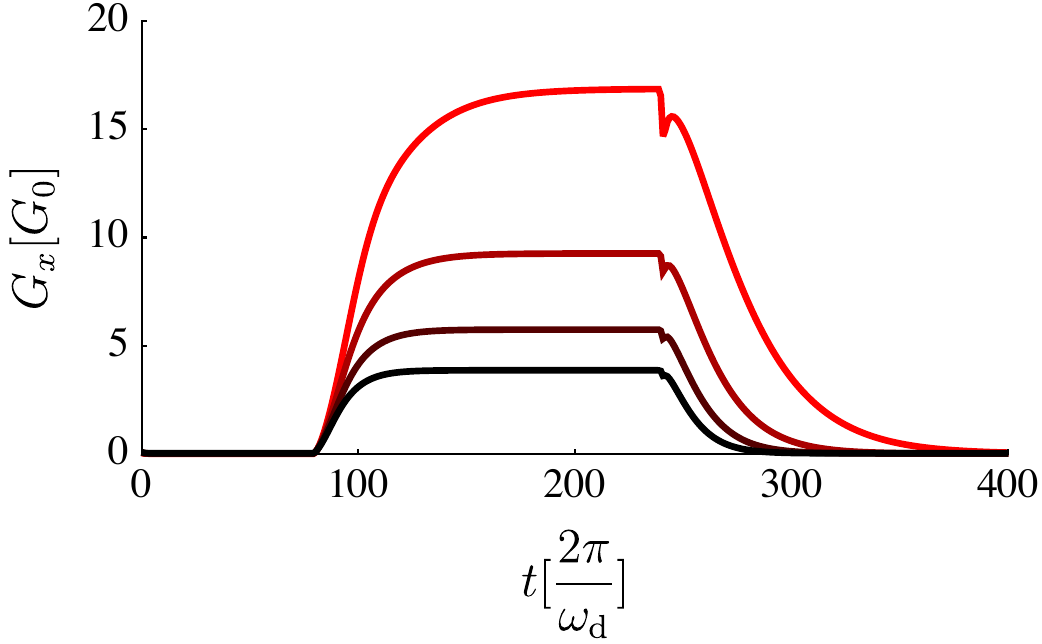}
    \caption{Longitudinal photoconductivity as a function of time for various dissipation strengths. The driving field is switched on after a time of $80$ driving periods and switched off after another $160$ driving periods. From top to bottom the dissipation rates increase as $\alpha=2,3,4,5$. We use $E_\mathrm{d}=5\mathrm{MVm^{-1}}$.
    }
    \label{switching}
\end{figure}

For the purpose of using the photoconductivity in coherently driven nanoribbons as a photoswitch, we present a protocol in which the transversal driving field at the optimal frequency $\omega^*_d$ is turned on and off on a time scale of around $1\si{\pico\second}$. The longitudinal photoconductivity is shown in Fig.~\ref{switching} for different dissipation rates as a function of time. 
In general, after the driving field is turned on, the current saturates to its steady-state value on the scale of picoseconds. When the graphene nanoribbon is no longer illuminated, the current decays to zero again. Both, the process of building up the current and the relaxation back to zero, are determined by the intrinsic dissipative processes and therefore scale as $\propto 1/\gamma$. Thus, strong dissipation implies that the steady state is reached quickly such that the photocurrent saturates earlier (black curve) in comparison to the case of low dissipation (red curve) where the current continues to grow during the protocol. We also note that the steady-state value of the photoconductivity decreases with increasing dissipation.
For the use of this system as an optical switch, not only a sufficiently large difference between the current for the non-driven and driven regime is desirable, but also a stabilized current response in both cases. 
In this sense there is a compromise between the operation time and the different dissipative channels present in the device.
For our choice of dissipation coefficients, similar to those used to describe previous experiments in graphene \cite{Nuske20}, the switching times are on the order of picoseconds which is equivalent to a clock-rate of about 1 THz.

\begin{figure}[tb]
    \centering 
    \includegraphics[width=1.0\linewidth]{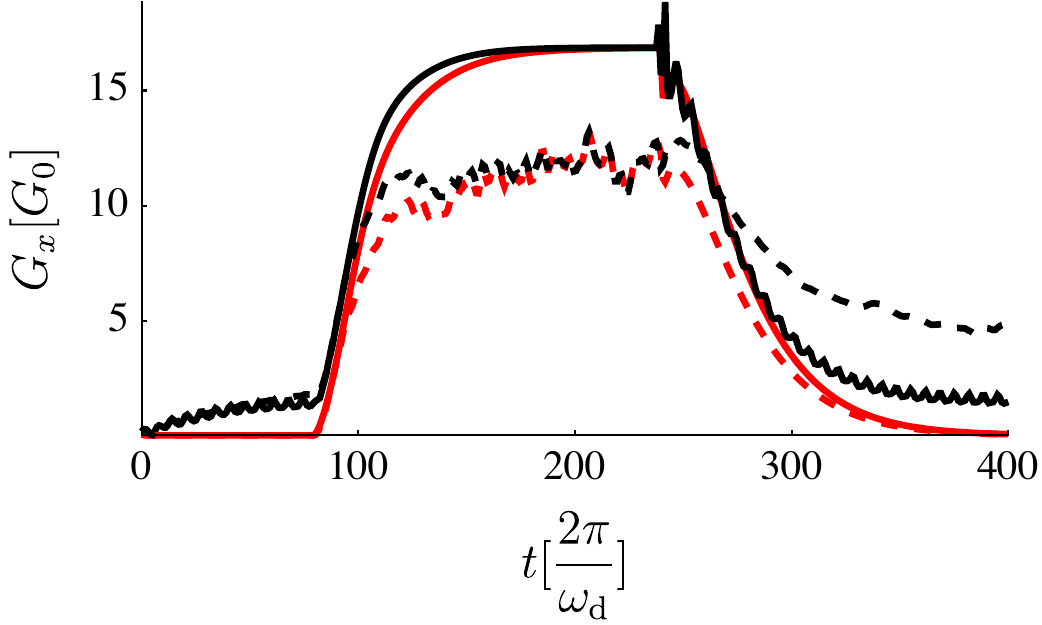}
    \caption{Time-dependence of photoconductivity using two different protocols. First, the ampltiude of the drive is switched on and off while keeping the drive frequency constant (red lines). Second, the drive frequency is modulated around the bare electronic band gap (black lines). Solid lines indicate the photoconductivity obtained with coherent driving, whereas dashed lines correspond to the case of noisy driving. We use $\alpha=2$ and $E_\mathrm{d}=5\mathrm{MVm^{-1}}$.
    }
    \label{freqmodulation}
\end{figure}

In Fig.~\ref{freqmodulation} we compare the switching dynamics of the photoconductivity for coherent (red line) and noisy (red dashed line) driving. We also show, with black lines, the photoconductivity dynamics for an alternative protocol in which, instead of modulating the amplitude of the drive, we modulate the drive frequency around $\Delta$.
In the later case, we switch the driving frequency from an initial value $\hbar\omega^i_d<\Delta$ to the optimal value $\omega^f_d=\omega^*_d$ and back to $\omega^i_d$ while keeping the driving field strength $E_d$ fixed.
In both cases, we find that the initial and final stage of the switching behavior is very similar with and without noisy driving. However, for the frequency switching protocol, in the presence of noisy driving, the switch-off process is slower indicating that it is difficult to the system to release the excess of energy back to the bath.
 We emphasize that in both protocols coherent excitation improves the performance of the switching behavior by displaying a larger steady-state current. In the presence of incoherent driving (dashed lines) the photoconductivity is suppressed even under the action of the drive so it is more difficult to switch between the different regimes.

\begin{figure}
    \centering 
\includegraphics[width=0.96\linewidth]{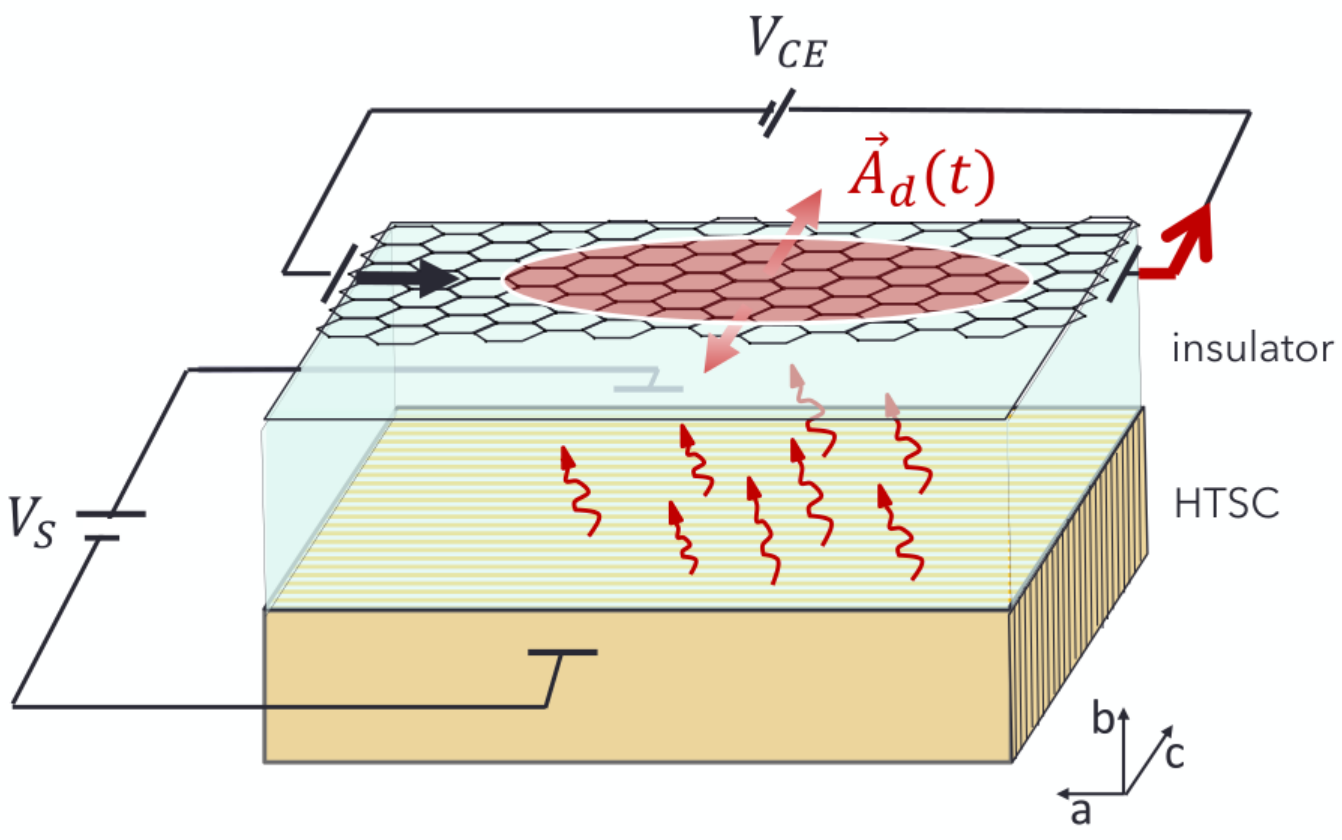}
    \caption{Transistor design, based on a heterostructure of a graphene nanoribbon, an insulating layer (transparent slab) and a layered high-$T_c$ superconductor (ochre), from top to bottom. By means of a switching voltage $V_S$, the high-$T_c$ superconductor (HTSC) emits light in the THz range that controls the current passing through the graphene nanoribbon. The latter is probed by applying a collector-emitter voltage $V_{CE}$.
    }
    \label{transistorprototype}
\end{figure}

We propose to use the light-mediated switching of conductivity presented here as a technology, in particular for graphene-based photoswitches or transistors without the use of external lasers. 
In Fig.~\ref{transistorprototype}, we show an example in which we propose to fabricate a heterostructure, composed of a graphene nanoribbon, and a high-$T_c$ superconductor. The two materials are separated by an insulating layer. Applying a DC voltage to the high-$T_c$ material along the c-axis ($V_S$ in the figure) induces voltage-generated light emission, due to the AC Josephson effect~\cite{JOSEPHSON1962251,Ulrich,Ozyuzer07}. The emitted light drives the nanoribbon, resulting in a coherently generated photoconductivity, as we described in the previous section. We note that this switching voltage pulse $V_S$ translates into a pulse protocol in which the frequency is tuned from small values to large values and back, similar to the frequency protocol described in the previous section. The heterostructure device that we describe here, constitutes a DC operated conductivity switch, in which the AC Josephson effect provides the frequency upconversion, and the coherent photoconductivity of a nanoribbon the frequency downconversion. The overall functionality is reminiscent of a transistor.

We note that light emission on the order of a few THz has been demonstrated in layered cuprate superconductors like LSCO or BSCCO.  Superconductors of the type La$_{2-x}$Ba$_x$CuO$_4$ can emit light with frequencies around $0.5$ THz \cite{Nicoletti22} whereas Bi$_2$Sr$_2$CaCu$_2$O$ _8$ reaches between $0.85$ and $2.5$ THz \cite{Ozyuzer07}. Even higher frequencies up to $11$ THz in Bi$_2$Sr$_2$CaCu$_2$O$_{8+\delta}$ have been experimentally achieved \cite{Borodianskyi} which matches the size-dependent electronic band gaps of armchair graphene nanoribbons.
We emphasize that in~\cite{Averitt} it was demonstrated that LSCO was cleaved along the c-axis, providing the type of sample that we propose to use. With these experimental achievements, our proposal constitutes a realistic device in the field of coherent electronics.



\section{Conclusion}
We have demonstrated coherently-enhanced photoconductivity of a graphene naoribbon. Specifically, we have shown that driving an armchair nanoribbon with coherent light results in a larger photoconductivity compared to driving with incoherent light. We utilize a master equation-based methodology to generate these predictions, and compare them to a standard semiclassical Boltzmann approach. We show a discrepancy of roughly 25$\%$ between the photoconductivity obtained using these two approaches, for realistic parameters. This deviation is due to purely coherent phenomena linked to Floquet physics that are not be captured by a quasiequilibrium rate equation formalism or Boltzmann-like treatments that do not take into account the coherence of the dynamics. We find that this discrepancy diminishes as we introduce incoherent driving, further supporting our understanding that this is a coherent current contribution. We propose to use this phenomenon of coherently enhanced photoconductivity, as part of a heterostructure device, composed of a nanoribbon and a high-$T_c$ superconductor. By applying a DC voltage to the high-$T_c$ superconductor, THz radiation is emitted due to the AC Josephson effect, which in turn drives the nanoribbon coherently. Thus, the DC voltage applied to the superconductor controls whether the nanoribbon is conducting or insulating, which constitutes functionality comparable to that of a transistor. With this we put forth a coherent electronic device within current experimental reach.

\section{Acknowledgments}
We acknowledge funding by the Deutsche Forschungsgemeinschaft (DFG, German Research Foundation) “SFB-925” Project No. 170620586 and the Cluster of Excellence “Advanced Imaging of Matter” (EXC 2056), Project No. 390715994. The project is co-financed by ERDF of the European Union and by ’Fonds of the Hamburg Ministry of Science, Research, Equalities and Districts (BWFGB)’.

\bibliography{references}

\end{document}